\documentclass[11pt]{article}

\usepackage[T1]{fontenc}
\usepackage[utf8]{inputenc}
\usepackage[a4paper,margin=2.6cm]{geometry}
\usepackage{amsmath,amssymb,mathtools,bm}
\usepackage{graphicx}
\usepackage{booktabs}
\usepackage{array}
\usepackage{microtype}
\usepackage[numbers,sort&compress]{natbib}
\usepackage{xcolor}
\usepackage{hyperref}

\newcommand{\Msun}{M_\odot}
\newcommand{\astar}{a_\star}
\newcommand{\Qtilde}{\widetilde{Q}}
\newcommand{\qdark}{q_{X}}
\newcommand{\muphi}{\mu_{\varphi}}
\newcommand{\Ecloud}{E_{\rm c}}
\newcommand{\dd}{\mathrm{d}}
\newcommand{\pos}[1]{\left[#1\right]_{+}}

\hypersetup{
  colorlinks=true,
  linkcolor=blue!55!black,
  citecolor=blue!55!black,
  urlcolor=blue!55!black,
  pdftitle={A Minimal Toy Model of Charged Superradiant Spin-Down in Extreme-Mass-Ratio Inspirals},
  pdfauthor={Bo Wang}
}

\title{A Minimal Toy Model of Charged Superradiant Spin-Down in Extreme-Mass-Ratio Inspirals}

\author{%
Bo Wang\\[3pt]
\small Department of Astronomy, School of Physical Sciences and\\[-1pt]
\small CAS Key Laboratory for Research in Galaxies and Cosmology,\\[-1pt]
\small School of Astronomy and Space Science, University of Science and Technology of China,\\[-1pt]
\small Hefei, 230026, China.\\[4pt]
\small Correspondence: \href{mailto:ymwangbo@ustc.edu.cn}{ymwangbo@ustc.edu.cn}}
\date{}

\begin{document}

\maketitle

\begin{abstract}
Long-lived extreme-mass-ratio inspirals (EMRIs) are sensitive to slow changes of the mass and spin of their central black holes. We construct a deliberately minimal toy model in which an effective hidden charge opens a charged-superradiance channel, a bosonic cloud grows from a seed, and energy and angular momentum are transferred from a rotating black hole to the cloud. The model is not a Kerr--Newman perturbation calculation. Instead, it consists of a threshold function, one phenomenological rate coefficient, and balance equations that conserve the combined black-hole--cloud energy. This reduction yields an analytic relation between mass loss and spin-down and makes the roles of the growth coefficient, seed energy, and superradiant margin transparent.

For an illustrative system with initial mass $2\times10^{6}\,\Msun$, spin $0.90$, effective charge-to-mass ratio $0.30$, and a one-year evolution time, the chosen benchmark reaches $M/M_{0}=0.8933$ and $\astar=0.7704$. These values are then passed to \textsc{FastEMRIWaveforms} as two stationary Kerr snapshots. The resulting figures show that the endpoint state has a shifted time-domain oscillation pattern and a lower-frequency harmonic structure relative to the initial state. Because the snapshots are regenerated independently, they do not constitute a phase-connected, time-dependent waveform and cannot by themselves support a detectability claim. The value of the construction is instead diagnostic: it isolates a compact set of assumptions, exposes parameter degeneracies, and identifies the minimum ingredients required for a future self-consistent calculation. In this sense, the model provides a reproducible bridge between charged-superradiance intuition and EMRI waveform studies without presenting its benchmark as an astrophysical forecast.
\end{abstract}

\noindent\textbf{Keywords:} extreme-mass-ratio inspirals; black-hole superradiance; ultralight bosons; hidden charge; toy model; gravitational waves

\section{Introduction}
\label{sec:introduction}

Extreme-mass-ratio inspirals (EMRIs) contain a stellar-mass compact object orbiting a massive black hole with mass ratio $\mu/M\ll1$. Their signals can remain in the millihertz band for $10^{4}$--$10^{5}$ orbital cycles, making phase evolution a sensitive probe of the central spacetime and of slowly varying environmental effects \citep{AmaroSeoane:2017,Gair:2013,GairEtAl:2004,Hughes:2003,BarackCutler:2004,DrascoHughes:2006,BarackPound:2019}. LISA is the reference detector in this paper, although the underlying question is also relevant to other planned space missions, including TianQin and Taiji \citep{Luo:2016TianQin,FanEtAl:2020TianQinEMRI,LuoEtAl:2026TianQinReview,HuWu:2017Taiji,CaiEtAl:2017GWPhysics}.

Ultralight bosonic fields are widely studied as dark-sector and dark-matter candidates with cosmological and astrophysical implications \citep{Marsh:2016,HuiOstrikerTremaineWitten:2017}. Around a rotating black hole, a bosonic quasi-bound state satisfying a superradiant condition can extract energy and angular momentum \citep{Zeldovich:1972,Starobinsky:1973,PressTeukolsky:1972,ZourosEardley:1979,Detweiler:1980,CardosoDiasLemosYoshida:2004,Dolan:2007,Brito:2015}. This process can build a macroscopic cloud, including stationary threshold configurations and corresponding scalar-hair solutions \citep{ArvanitakiDubovsky:2011,ArvanitakiBaryakhtarHuang:2015,Hod:2012,HerdeiroRadu:2014}. If the field carries a charge under a hidden $U(1)$, the threshold includes an electrostatic term, as in the charged-scalar problem on a Kerr--Newman background \citep{Bekenstein1998charged,FuruhashiNambu:2004}. The hidden-sector interpretation is useful because the effective charge is not an ordinary electromagnetic charge that must survive neutralization by an astrophysical plasma.

A precision prediction for this system would require at least four coupled ingredients: a charged quasi-bound-state spectrum, conservation of mass, angular momentum, and hidden charge, the cloud stress-energy and its interaction with the secondary, and a waveform model on the resulting time-dependent nonvacuum spacetime. Those ingredients are not presently combined in the calculation underlying this manuscript. Treating the reduced model as though it were such a prediction would obscure rather than clarify what has been learned.

We therefore adopt the opposite strategy and formulate the calculation explicitly as a toy model. The purpose of a toy model is not to approximate every relevant interaction. It is to retain the smallest set of rules needed to answer a sharply limited question:
\begin{quote}
If a phenomenological charged-superradiance channel changes a massive black hole's mass and spin over an EMRI observation, how do stationary waveform snapshots at the beginning and end of that evolution differ?
\end{quote}
This question can be answered while keeping all assumptions visible. It also separates a robust kinematic statement---different central parameters generate different EMRI frequencies---from the stronger and presently unsupported claim that the displayed curves form a detector-ready signal.

The paper is organized as follows. Section~\ref{sec:scope} defines the model's scope. Section~\ref{sec:model} gives the threshold, growth law, and balance equations and derives an analytic mass--spin relation. Section~\ref{sec:scales} identifies the dimensionless control parameters and the transient window. Section~\ref{sec:waveform} defines the stationary-snapshot waveform diagnostic. Section~\ref{sec:benchmark} presents the benchmark results. Sections~\ref{sec:lessons} and~\ref{sec:limitations} discuss what can and cannot be inferred, and Sec.~\ref{sec:conclusion} concludes. We use $G=c=1$ except when converting to physical units.

\section{Scope and closure assumptions}
\label{sec:scope}

The toy model has three dynamical variables,
\begin{equation}
  y(t)=\{M(t),J(t),\Ecloud(t)\},
  \label{eq:state}
\end{equation}
where $M$ and $J$ are the black-hole mass and angular momentum and $\Ecloud$ is the cloud energy. The dimensionless spin is $\astar=J/M^{2}$. Four closure assumptions complete the model.

\begin{enumerate}
  \item \textbf{Effective charged rotating background.} The parameters $\astar$ and $\Qtilde$ enter horizon-inspired functions. The charge ratio $\Qtilde$ is held fixed while $M$ changes. This is a phenomenological closure, not hidden-charge conservation.
  \item \textbf{One bosonic mode.} A single mode with azimuthal number $m_{\rm b}=1$ and principal number $n=2$ carries the extracted energy and angular momentum.
  \item \textbf{Margin-controlled growth.} The instability rate is proportional to the positive part of a dimensionless superradiant margin. All unresolved microphysics is compressed into one coefficient $\kappa$.
  \item \textbf{One-way waveform map.} The black-hole evolution supplies endpoint values of $M$ and $\astar$ to a vacuum Kerr waveform generator. The orbit and waveform do not feed back on the cloud evolution.
\end{enumerate}

These assumptions intentionally exclude cloud self-gravity, scalar drag, mode mixing, hidden-charge flux, cloud gravitational radiation, and a time-dependent metric \citep{EastPretorius:2017,BaryakhtarLasenbyTeo:2017,BritoEtAl:2017PRL,East:2018}. They also distinguish the present calculation from studies of scalar-field effects in EMRIs, stationary scalar environments, or explicit environmental forces \citep{GuoEtAl:2022ScalarEMRI,Maselli:2020,ZhangYang:2020,BaumannBertoneStoutTomaselli:2022,VicenteCardoso:2022,Brito:2023ShaSr,Dyson:2025,Polcar:2025}.

\section{The toy black-hole--cloud model}
\label{sec:model}

\subsection{Horizon-inspired threshold functions}

Define
\begin{equation}
  s(\astar,\Qtilde)=\sqrt{1-\astar^{2}-\Qtilde^{2}},
  \qquad
  \rho_{+}=1+s,
  \label{eq:horizonproxy}
\end{equation}
with the domain $\astar^{2}+\Qtilde^{2}<1$. The numerical implementation used to make Fig.~\ref{fig:spin_evolution} employs the dimensionless proxies
\begin{equation}
  \overline{\Omega}_{\rm H}^{\rm toy}
  =\frac{\astar}{2\rho_{+}},
  \qquad
  \overline{\Phi}_{\rm H}^{\rm toy}
  =\frac{\Qtilde\rho_{+}}{\rho_{+}^{2}+\astar^{2}}.
  \label{eq:horizonfunctions}
\end{equation}
The second expression has the form of the Kerr--Newman horizon potential. The first is a Kerr-like frequency proxy with the charged discriminant inserted. For an exact Kerr--Newman geometry one would instead use $M\Omega_{\rm H}=\astar/(\rho_{+}^{2}+\astar^{2})$. Equation~\eqref{eq:horizonfunctions} is the prescription used for the results shown below; making this approximation explicit is part of the toy-model interpretation.

The bound-state frequency is represented by the leading hydrogenic expression
\begin{equation}
  \overline{\omega}_{R}
  =\alpha_{\rm eff}
  \left(1-\frac{\alpha_{\rm eff}^{2}}{2n^{2}}\right),
  \label{eq:hydrogenic}
\end{equation}
where $\alpha_{\rm eff}$ is treated as a constant input. We then define the dimensionless superradiant margin
\begin{equation}
  \delta_{\rm sr}(\astar)
  =m_{\rm b}\overline{\Omega}_{\rm H}^{\rm toy}
  +\qdark\overline{\Phi}_{\rm H}^{\rm toy}
  -\overline{\omega}_{R}.
  \label{eq:margin}
\end{equation}
The switch is open only when $\delta_{\rm sr}>0$. This mirrors the structure of the charged-superradiance condition $\omega_{R}<m\Omega_{\rm H}+q\Phi_{\rm H}$ without claiming that $\qdark$ is a convention-independent microscopic charge.

\subsection{Growth law and balance equations}

The physical growth rate is defined by
\begin{equation}
  \Gamma(t)=\frac{\kappa}{M(t)}\pos{\delta_{\rm sr}(\astar(t))},
  \qquad
  \pos{x}=\max(x,0),
  \label{eq:growth}
\end{equation}
and the cloud obeys
\begin{equation}
  \frac{\dd\Ecloud}{\dd t}=2\Gamma\Ecloud.
  \label{eq:cloudgrowth}
\end{equation}
The coefficient $\kappa$ is dimensionless and phenomenological. It should not be identified with a computed Kerr--Newman eigenvalue.

Energy and angular momentum are transferred according to
\begin{align}
  \frac{\dd M}{\dd t}
  &=-\frac{\dd\Ecloud}{\dd t},
  \label{eq:massbalance}\\
  \frac{\dd J}{\dd t}
  &=-\frac{m_{\rm b}}{\omega_{R}(t)}
  \frac{\dd\Ecloud}{\dd t},
  \qquad
  \omega_{R}(t)=\frac{\overline{\omega}_{R}}{M(t)}.
  \label{eq:angularbalance}
\end{align}
Growth is stopped if the superradiant margin closes, the horizon domain is approached, or a numerical safety cap $\Ecloud=f_{\rm cap}M$ is reached. The cap is not active for the benchmark endpoint reported below.

Equations~\eqref{eq:cloudgrowth} and~\eqref{eq:massbalance} immediately give the invariant
\begin{equation}
  M(t)+\Ecloud(t)=M_{0}+\Ecloud(0).
  \label{eq:energyinvariant}
\end{equation}
This is the central conservation relation of the reduced evolution. It applies within the toy model and does not imply that radiation and other channels vanish in the physical system.

\subsection{Analytic mass--spin trajectory}
\label{sec:analytic}

The balance equations contain more information than a numerical integration alone makes apparent. Eliminating $\Ecloud$ and $t$ gives
\begin{equation}
  \frac{\dd J}{\dd M}
  =\frac{m_{\rm b}M}{\overline{\omega}_{R}}.
  \label{eq:dJdM}
\end{equation}
For constant $\overline{\omega}_{R}$, integration yields
\begin{equation}
  J(M)=J_{0}
  +\frac{m_{\rm b}}{2\overline{\omega}_{R}}
  \left(M^{2}-M_{0}^{2}\right).
  \label{eq:JofM}
\end{equation}
Writing $x=M/M_{0}$ and $J_{0}=a_{\star 0}M_{0}^{2}$, the spin follows the closed-form trajectory
\begin{equation}
  \astar(x)
  =\frac{a_{\star 0}}{x^{2}}
  +\frac{m_{\rm b}}{2\overline{\omega}_{R}}
  \left(1-\frac{1}{x^{2}}\right).
  \label{eq:spinmassrelation}
\end{equation}
Thus $\kappa$ does not determine the path in the $M$--$\astar$ plane; it determines how quickly that path is traversed. Equation~\eqref{eq:spinmassrelation} also reproduces the benchmark endpoint: $x=0.8933$, $a_{\star 0}=0.90$, $m_{\rm b}=1$, and $\overline{\omega}_{R}=0.3542$ give $\astar\simeq0.770$.
This analytic reduction is the main benefit of the toy-model formulation. It exposes which result follows from balance alone and which result depends on the chosen growth law.

\section{Control parameters and the transient window}
\label{sec:scales}

In a slowly varying background, the hierarchy required for an adiabatic EMRI diagnostic is
\begin{equation}
  \tau_{\rm orb}\ll \tau_{\rm bg},\tau_{\rm RR},
  \label{eq:adiabatic}
\end{equation}
where $\tau_{\rm bg}\sim\Gamma^{-1}$ is the background-evolution time and $\tau_{\rm RR}$ is the orbital radiation-reaction time. A time-resolved effect is largest when $\tau_{\rm bg}$ is also comparable to the observing time. This motivates the window
\begin{equation}
  \tau_{\rm orb}\ll
  \tau_{\rm bg}\sim\tau_{\rm obs},
  \label{eq:transientwindow}
\end{equation}
but the benchmark is chosen to illustrate this window rather than derived from an astrophysical population.

At early times, when $M$ and $\delta_{\rm sr}$ can be approximated as constant, Eq.~\eqref{eq:cloudgrowth} gives
\begin{equation}
  \Ecloud(t)\simeq\Ecloud(0)
  \exp\!\left[\frac{2\kappa\delta_{\rm sr,0}}{M_{0}}t\right].
  \label{eq:earlygrowth}
\end{equation}
The time required to reach a chosen cloud fraction $f$ is therefore
\begin{equation}
  \frac{t_{f}}{M_{0}}
  \simeq
  \frac{1}{2\kappa\delta_{\rm sr,0}}
  \ln\!\left(\frac{f}{\Ecloud(0)/M_{0}}\right).
  \label{eq:turnovertime}
\end{equation}
Only the combination $2\kappa\delta_{\rm sr,0}$ fixes the exponential clock, while the seed enters logarithmically. Consequently, $\kappa$ and $\Ecloud(0)$ are strongly degenerate: changing either one mainly translates the rapid-growth episode along the time axis. This degeneracy is more informative than assigning a unique physical meaning to the benchmark value of $\kappa$.

For the inputs in Table~\ref{tab:benchmark}, $\delta_{\rm sr,0}\simeq0.097$. The cloud-energy e-folding time is approximately $M_{0}/(2\kappa\delta_{\rm sr,0})$, or about 12 days for $M_{0}=2\times10^{6}\,\Msun$. Starting from a $10^{-12}M_{0}$ seed then delays the visible turnover until late in the one-year interval, consistent with Fig.~\ref{fig:spin_evolution}.

\section{Stationary-snapshot waveform diagnostic}
\label{sec:waveform}

The waveform stage uses the public \textsc{FastEMRIWaveforms} package \citep{ChuaKatz:2021FASTEMRI,KatzChua:2021FASTEMRI}. The selected model, \texttt{FastKerrEccentricEquatorialFlux}, is a vacuum Kerr model. Hidden charge and the cloud do not enter its field equations. Their only role is to supply two sets of central parameters:
\begin{equation}
  \mathcal{S}_{\rm i}=(M_{0},a_{\star 0}),
  \qquad
  \mathcal{S}_{\rm f}=(M_{\rm f},a_{\star{\rm f}}).
  \label{eq:snapshots}
\end{equation}

For a generic bound Kerr orbit, the harmonic frequencies have the schematic form
\begin{equation}
  \omega_{knm}
  =k\Omega_{r}+n\Omega_{\theta}+m\Omega_{\phi}.
  \label{eq:harmonics}
\end{equation}
Changing $M$ and $\astar$ therefore moves many harmonics coherently. A simple circular-equatorial estimate is
\begin{equation}
  \Omega_{\phi}
  =\frac{1}{M\left[{\left(r/M\right)}^{3/2}+\astar\right]},
  \label{eq:orbitalfrequency}
\end{equation}
which shows directly that both the mass rescaling and the spin change can shift the instantaneous frequency \citep{Bardeen:1973}.

The calculation is a \emph{snapshot comparison}, not an evolving waveform. Each one-hour segment is computed independently at its assigned stationary state, with the initial orbital parameter specified separately for that segment. Likewise, the spectra compare stationary initial and final states; they do not Fourier transform a continuously evolved one-year signal. It is therefore meaningful to discuss endpoint frequency and morphology changes, but not an accumulated phase lag between the displayed curves.
This distinction also limits detector statements. A LISA sensitivity curve is plotted in Fig.~\ref{fig:spectrum} only as a frequency-band reference. No consistent detector response, time-delay interferometry, noise-weighted inner product, signal-to-noise ratio, mismatch, or parameter-estimation calculation is performed.

\section{Benchmark illustration}
\label{sec:benchmark}

\subsection{Inputs}

Table~\ref{tab:benchmark} records the parameters used in the numerical evolution and waveform snapshots. Unlike a parameter-estimation benchmark, these values are selected to place the rapid part of the toy evolution near the end of a one-year interval. The mode-frequency input $\alpha_{\rm eff}=0.36$ corresponds to $\overline{\omega}_{R}=0.3542$ through Eq.~\eqref{eq:hydrogenic}. If interpreted as $M_{0}\muphi$, it is associated with a boson rest energy of order $2.4\times10^{-17}\,\mathrm{eV}$ for the chosen mass, but this conversion should not be overinterpreted because $\alpha_{\rm eff}$ is held fixed as $M$ evolves.

\begin{table}[t]
\centering
\caption{Inputs for the illustrative toy evolution and stationary waveform snapshots. The rate, seed, and cap are phenomenological choices rather than fitted physical parameters.}
\label{tab:benchmark}
\begin{tabular}{>{\raggedright\arraybackslash}p{0.49\textwidth}p{0.30\textwidth}}
\toprule
Quantity & Value \\
\midrule
Initial primary mass $M_{0}$ & $2.0\times10^{6}\,\Msun$ \\
Secondary mass $\mu$ & $10\,\Msun$ \\
Initial spin $a_{\star 0}$ & $0.90$ \\
Effective charge ratio $\Qtilde$ & $0.30$ \\
Mode numbers $(m_{\rm b},n)$ & $(1,2)$ \\
Frequency parameter $\alpha_{\rm eff}$ & $0.36$ \\
Effective hidden coupling $\qdark$ & $0.70$ \\
Growth coefficient $\kappa$ & $5.0\times10^{-5}$ \\
Initial cloud fraction $\Ecloud(0)/M_{0}$ & $10^{-12}$ \\
Safety cap $f_{\rm cap}$ & $0.50$ \\
Evolution interval & $1~\mathrm{yr}$ \\
Snapshot eccentricity and inclination parameter & $e_{0}=0.10,\ x_{0}=1$ \\
Reference luminosity distance & $1~\mathrm{Gpc}$ \\
\bottomrule
\end{tabular}
\end{table}

\subsection{Toy mass and spin evolution}

\begin{figure}[t]
\centering
\includegraphics[width=0.86\textwidth]{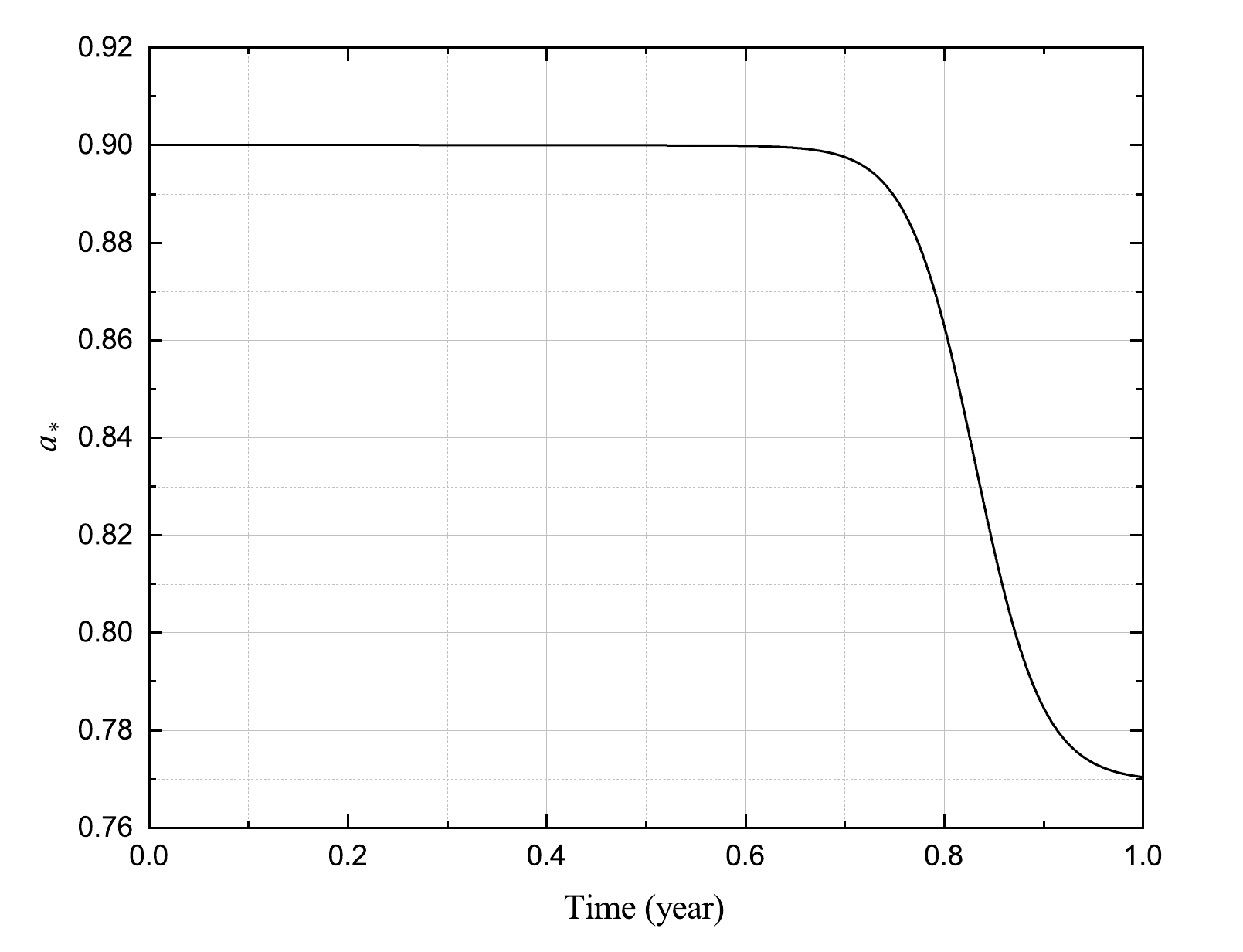}
\caption{Toy black-hole--cloud evolution. The initial seed delays visible evolution until late in the one-year interval. The black-hole spin falls from $0.90$ to $0.7704$, while the mass falls to $0.8933M_{0}$. The curve illustrates the chosen phenomenological clock; it is not a numerical Kerr--Newman superradiance solution.}
\label{fig:spin_evolution}
\end{figure}

Figure~\ref{fig:spin_evolution} shows the numerical evolution obtained from Eqs.~\eqref{eq:growth}--\eqref{eq:angularbalance}. The long plateau is a direct consequence of the tiny seed. Once exponential growth becomes visible, the mass and spin follow the analytic relation in Eq.~\eqref{eq:spinmassrelation}. At one year the numerical endpoint is
\begin{equation}
  \frac{M_{\rm f}}{M_{0}}=0.893275,
  \qquad
  a_{\star{\rm f}}=0.7704.
  \label{eq:endpoint}
\end{equation}
The mass loss is $10.7\%$, so the $50\%$ safety cap does not determine the endpoint. The sharpness and timing of the transition do depend on the rate law and seed and are not presented as an astrophysical prediction.

\subsection{Time-domain endpoint comparison}

\begin{figure}[t]
\centering
\includegraphics[width=0.86\textwidth]{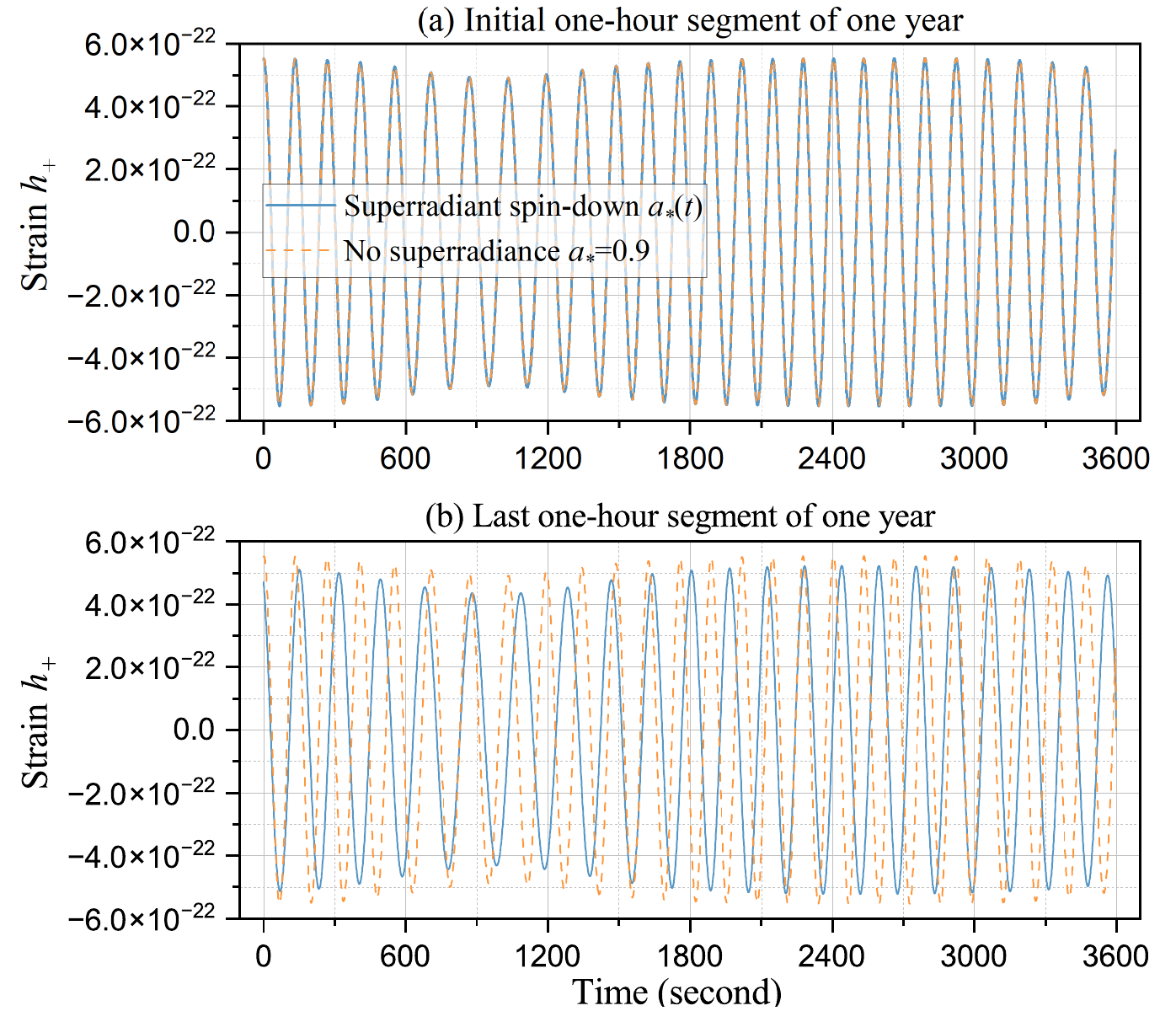}
\caption{Source-frame $h_{+}$ waveform snapshots. The upper panel shows the coincident waveforms at the common initial state. The lower panel compares an independently generated endpoint snapshot, using $M_{\rm f}=1.7866\times10^{6}\,\Msun$ and $a_{\star{\rm f}}=0.7704$, with a snapshot retaining $M_{0}=2\times10^{6}\,\Msun$ and $a_{\star 0}=0.90$. The segments are not phase connected across one year.}
\label{fig:waveforms}
\end{figure}

Figure~\ref{fig:waveforms} compares one-hour waveform snapshots. The two curves in the upper panel coincide because they are generated with the same initial state. In the lower panel, the blue curve uses the final values in Eq.~\eqref{eq:endpoint}, while the orange dashed curve retains the initial mass and spin. The distinct oscillation rates demonstrate that the endpoint substitution is visible at the waveform level.

The lower panel should not be read as the last hour of two waveforms that shared a common phase history for the preceding year. The independent regeneration of the segments removes that phase connection. Accordingly, the figure supports a statement about instantaneous endpoint morphology, not about accumulated dephasing.

\subsection{Frequency-domain endpoint comparison}

\begin{figure}[t]
\centering
\includegraphics[width=0.78\textwidth]{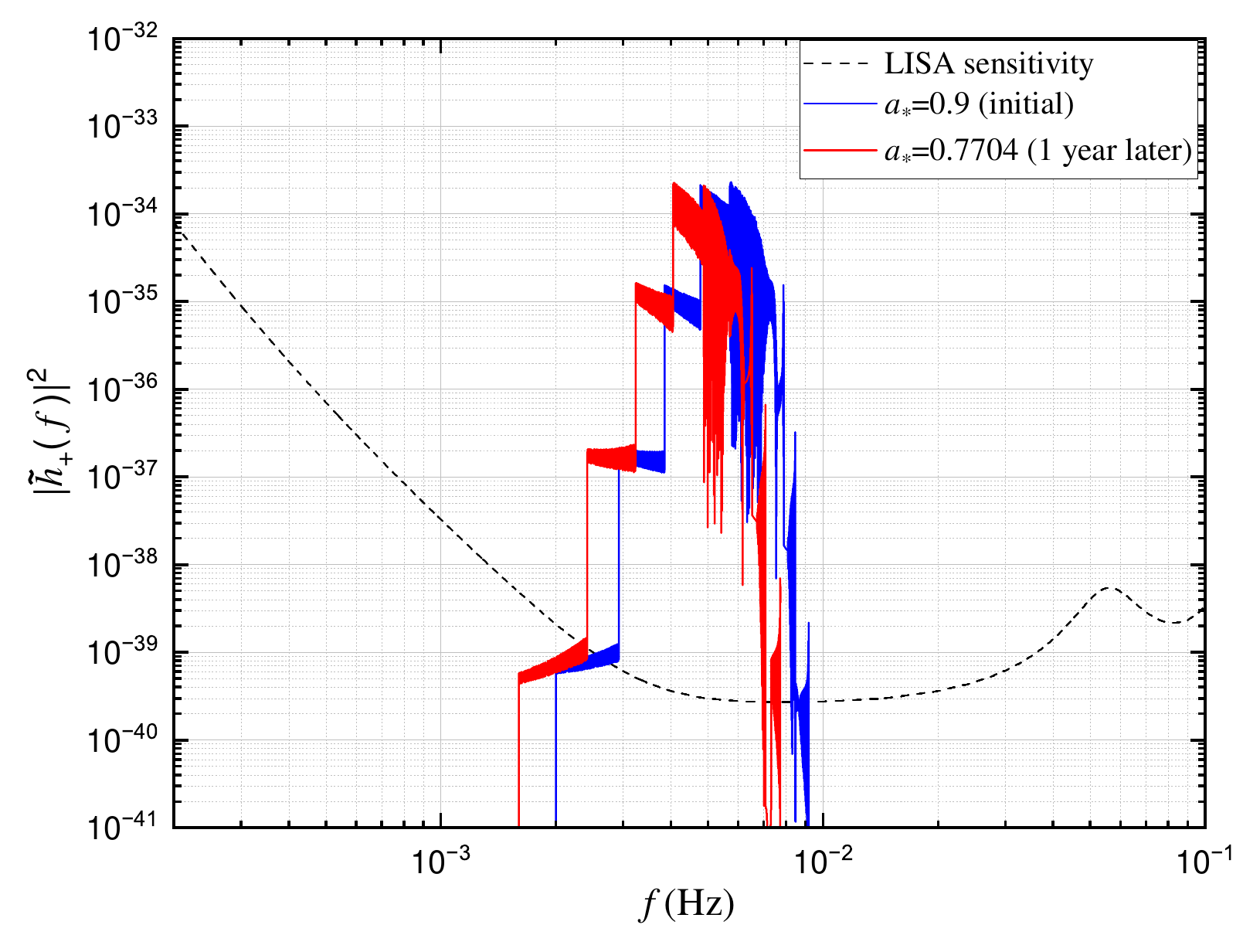}
\caption{Frequency-domain comparison of stationary initial and final snapshots. The final-state harmonic structure (red) is shifted to lower frequencies relative to the initial state (blue). The dashed LISA curve is shown only to indicate the relevant band; no detectability statistic is inferred.}
\label{fig:spectrum}
\end{figure}

Figure~\ref{fig:spectrum} gives the corresponding stationary frequency-domain diagnostic. The final-state harmonic support is displaced toward lower frequencies relative to the initial state. Because both $M$ and $\astar$ change, the plot does not isolate a pure spin effect. It is best understood as a visualization of the map
\begin{equation}
  (M_{0},a_{\star 0})
  \longrightarrow
  (M_{\rm f},a_{\star{\rm f}})
  \longrightarrow
  \widetilde{h}_{+}(f;M_{\rm f},a_{\star{\rm f}}).
  \label{eq:diagnosticmap}
\end{equation}
The dashed LISA curve locates the millihertz band but is not used to calculate a signal-to-noise ratio. In particular, the plotted waveform power and sensitivity curve do not constitute a complete noise-weighted comparison because the precise Fourier and response conventions are not specified.

\section{What the toy model teaches}
\label{sec:lessons}

\subsection{The endpoint trajectory is simpler than the time history}

Equation~\eqref{eq:spinmassrelation} shows that the benchmark's mass and spin changes are not independent outputs. Once the mode prescription is fixed, balance determines a one-dimensional trajectory in the $M$--$\astar$ plane. The rate model chooses only the position on that trajectory at a given time. This separation provides a useful criterion for more detailed calculations: the balance trajectory and the microphysical time evolution can be assessed independently.

\subsection{The onset time is not a unique measurement of the rate}

Equation~\eqref{eq:turnovertime} makes the $\kappa$--seed degeneracy explicit. An earlier onset can result from a larger growth coefficient, a larger initial cloud, or a larger initial superradiant margin. Without an independent model of the seed and charged spectrum, the late transition in Fig.~\ref{fig:spin_evolution} cannot be inverted into a unique instability rate.

\subsection{A frequency shift is weaker evidence than a phase-coherent residual}

The endpoint plots demonstrate a kinematic response to changed $M$ and $\astar$. The same response could be mimicked by choosing different constant source parameters. A genuinely discriminating observable would require one phase-connected waveform $h(t)$ generated while the background evolves, followed by a fit against optimized constant-background templates. Only the residual after such optimization could be compared with environmental effects such as disk torques, dark-matter drag, or third-body perturbations \citep{Kocsis:2011,Yunes:2011,Derdzinski:2019,Speri:2023,GuptaBongaChua:2021,Silva:2022}.

\subsection{The minimal next model is well defined}

The toy construction identifies a concrete sequence of refinements. Each item removes a specific closure assumption, allowing the refinements to be incorporated separately rather than all at once:
\begin{enumerate}
  \item replace Eq.~\eqref{eq:horizonfunctions} and the constant $\overline{\omega}_{R}$ with a numerical charged quasi-bound-state spectrum;
  \item evolve the hidden charge together with $M$, $J$, and $\Ecloud$;
  \item generate one orbital trajectory on the resulting slowly varying background rather than independent snapshots;
  \item include the cloud's gravitational field and direct force on the secondary;
  \item apply a full LISA response and compare optimized models using a noise-weighted likelihood.
\end{enumerate}

\section{Limitations and domain of validity}
\label{sec:limitations}

The model is intended for qualitative exploration under the hierarchy in Eq.~\eqref{eq:adiabatic}, and its limitations are substantial. At the model level, the rate law is a threshold ansatz: charged superradiance depends nontrivially on the mode quantum numbers, field mass and charge, black-hole parameters, and boundary conditions \citep{FuruhashiNambu:2004}, so the linear proportionality in Eq.~\eqref{eq:growth} is not a substitute for that spectrum. In addition, $\Qtilde$ and $\alpha_{\rm eff}$ remain fixed as $M$ changes. If they were constructed from fixed microscopic quantities, both would generally evolve. Holding them fixed is a convenient way to close the toy system, but it changes the threshold history.

At the benchmark and waveform level, more than ten percent of the black-hole mass is removed. Such a large change is useful for making the endpoint response visually clear, but it also makes neglected effects harder to justify; the result should therefore be viewed as an illustrative limiting case, not a typical event. Moreover, the waveform generator assumes stationary vacuum Kerr snapshots. It neither propagates an EMRI through the toy time-dependent background nor represents a charged or cloud-deformed geometry, and no phase-continuous residual, mismatch, or parameter bias is calculated.

Finally, the plotted LISA sensitivity is schematic in this context. Mission forecasts require the detector response, observation window, noise model, source orientation, and consistent Fourier normalization \citep{Cutler:1998,CornishRubbo:2003,RobsonCornishLiu:2019}, none of which is supplied by the visual overlap alone. These limitations are not peripheral caveats; they define the boundary of every conclusion below.

\section{Conclusions}
\label{sec:conclusion}

We have recast a charged-superradiance EMRI calculation as an explicit toy model. The reformulation leads to four conclusions.

\begin{enumerate}
  \item A threshold-controlled growth law plus energy and angular-momentum balance is sufficient to generate a transparent, reproducible mass--spin history, but the growth coefficient remains phenomenological.
  \item The balance equations imply the analytic mass--spin relation in Eq.~\eqref{eq:spinmassrelation}. The rate and seed control when the system moves along this trajectory, not the trajectory itself.
  \item For the deliberately strong benchmark, the one-year endpoint is $M/M_{0}=0.8933$ and $\astar=0.7704$. Stationary EMRI snapshots at this endpoint exhibit altered oscillation patterns and a lower-frequency harmonic structure relative to the initial state.
  \item The figures do not form a phase-connected evolving waveform and do not establish detectability or distinguishability. They identify an endpoint effect that a more complete model must track continuously.
\end{enumerate}

The appropriate role of this construction is therefore diagnostic rather than predictive. It provides a compact baseline to which a numerical charged spectrum, hidden-charge conservation, cloud backreaction, phase-continuous EMRI evolution, and detector-level inference can each be added and evaluated separately.

\section*{Data Availability}

No observational or external datasets were used in this study. The numerical information supporting the reported results is contained within the article and its figures.

\section*{Code Availability}

The computational procedures and model specifications used to obtain the reported results are described in the article. The waveform snapshots were generated with the publicly available \textsc{FastEMRIWaveforms} package \citep{ChuaKatz:2021FASTEMRI,KatzChua:2021FASTEMRI}.

\section*{Conflict of Interest}

The author declares no conflict of interest.

\section*{Author Contributions}

Bo Wang is the sole author.

\section*{Funding}

National Key R\&D Program of China (2024YFC2207500 and 2021YFC2203100), National Natural Science Foundation of China (92476203), the 111 Project (B23042), and the USTC Tang Scholar program.

\section*{Acknowledgments}

We appreciate the Supercomputing Center of USTC for providing high-performance AI infrastructure services. Generative-AI large language models were used as auxiliary tools during the analysis and preparation of the manuscript. All AI-assisted outputs were critically reviewed and verified by the author, who takes full responsibility for the scientific content and conclusions of this work.

\bibliographystyle{unsrtnat}
\bibliography{ref_PDU_darkcharged_EMRI}

\end{document}